

\font\twelverm=cmr10 scaled 1200    \font\twelvei=cmmi10 scaled 1200
\font\twelvesy=cmsy10 scaled 1200   \font\twelveex=cmex10 scaled 1200
\font\twelvebf=cmbx10 scaled 1200   \font\twelvesl=cmsl10 scaled 1200
\font\twelvett=cmtt10 scaled 1200   \font\twelveit=cmti10 scaled 1200
\font\twelvesc=cmcsc10 scaled 1200
\skewchar\twelvei='177   \skewchar\twelvesy='60


\def\twelvepoint{\normalbaselineskip=12.4pt plus 0.1pt minus 0.1pt
  \abovedisplayskip 12.4pt plus 3pt minus 9pt
  \belowdisplayskip 12.4pt plus 3pt minus 9pt
  \abovedisplayshortskip 0pt plus 3pt
  \belowdisplayshortskip 7.2pt plus 3pt minus 4pt
  \smallskipamount=3.6pt plus1.2pt minus1.2pt
  \medskipamount=7.2pt plus2.4pt minus2.4pt
  \bigskipamount=14.4pt plus4.8pt minus4.8pt
  \def\rm{\fam0\twelverm}          \def\it{\fam\itfam\twelveit}%
  \def\sl{\fam\slfam\twelvesl}     \def\bf{\fam\bffam\twelvebf}%
  \def\mit{\fam 1}                 \def\cal{\fam 2}%
  \def\sc{\twelvesc}               \def\tt{\twelvett}
  \def\sf{\twelvesf}
  \textfont0=\twelverm   \scriptfont0=\tenrm   \scriptscriptfont0=\sevenrm
  \textfont1=\twelvei    \scriptfont1=\teni    \scriptscriptfont1=\seveni
  \textfont2=\twelvesy   \scriptfont2=\tensy   \scriptscriptfont2=\sevensy
  \textfont3=\twelveex   \scriptfont3=\twelveex  \scriptscriptfont3=\twelveex
  \textfont\itfam=\twelveit
  \textfont\slfam=\twelvesl
  \textfont\bffam=\twelvebf \scriptfont\bffam=\tenbf
  \scriptscriptfont\bffam=\sevenbf
  \normalbaselines\rm}


\def\doublespace{\baselineskip=\normalbaselineskip \multiply\baselineskip by
 2}
\parindent=40pt
\newcount\firstpageno
\firstpageno=2
\footline={\ifnum\pageno<\firstpageno{\hfil}\else{
  \hskip 3.19truein\twelverm\folio\hfill}\fi}

\def\ic{{\rm I}\hskip-.5em{\rm C}}
\def\one{1\hskip-.37em 1}
\hsize=6.3truein
\hoffset=0.1truein
\vsize=8.5truein
\voffset=0truein
\topskip=0truein
\parskip=\medskipamount
\twelvepoint
\centerline{\bf Coherent States for the Hydrogen Atom}

\centerline{\it by}
\centerline{John R. Klauder}
\centerline{\it Departments of Physics and Mathematics}
\centerline{\it University of Florida, Gainesville, FL  32611}
\doublespace

\centerline{\bf Abstract}

{\narrower The long-standing problem of finding coherent states for the (bound
state portion of the) hydrogen atom is positively resolved.  The states in
question: (i)~are normalized and parameterized continuously, (ii)~admit a
resolution of unity with a positive measure, and (iii)~enjoy the property that
the temporal evolution of any coherent state by the hydrogen atom Hamiltonian
remains a coherent state for all time.\smallskip}
\vskip .15truein

\noindent{\bf Harmonic-oscillator coherent states}

In modern terms, the distinguished set of states Schr\"odinger introduced for
 the
harmonic oscillator$^1$ are now commonly known as coherent states$^2$ and are
given, for all $z\in \ic$, by
$$|z\rangle \equiv e^{(za^\dagger - z^*a)}|0\rangle = e^{-{1\over 2}|z|^2}
\sum_{n=0}^\infty {z^n\over \sqrt {n!}} |n\rangle, \eqno(1)$$
where $[a,a^\dagger]=1$ as usual.  Here $|n\rangle,n=0,1,2,\ldots$, denote
normalized eigenstates of the number operator $N$, $N|n\rangle=n|n\rangle$,
 which
may be identified with the eigenstates of the harmonic oscillator Hamiltonian
${\cal H}_0= \omega N=\omega a^\dagger a~(\hbar=1)$.  As such it follows that
$$e^{-i{\cal H}_0t} |z\rangle =e^{-{1\over 2}|z|^2}\sum_{n=0}^\infty
{z^ne^{-in\omega t}\over \sqrt{n!}} |n\rangle = |e^{-i\omega t} z\rangle
\eqno(2)$$
illustrating the fact that the time evolution of any such coherent state
 remains
within the family of coherent states; we shall refer to the property embodied
 in
(2) as temporal stability of the coherent states under ${\cal H}_0$, or, more
briefly, just as {\it temporal stability}.  Furthermore, these states are
evidently continuous in their label $z=x+iy$ and admit a resolution of unity
given by
$$\one = \int |z\rangle\langle z | dxdy/\pi\eqno(3)$$
integrated over $\ic$.  Continuity in the labels plus a resolution of unity
establish that the set $\{|z\rangle\}$ is a set of coherent states in the
 modern
sense of the term.$^3$
\vskip .15truein
\noindent{\bf Giving up the group}

Of course, there exist many other sets of states, which we also refer to as
coherent states, that are continuous in their labels and admit a resolution
of unity.  Such states may, for example, take the form, for all $\zeta\in
\ic$, given by
$$|\zeta\rangle = M(|\zeta|^2) \sum_{n=0}^\infty {\zeta^n\over\sqrt {\rho_n}}
|n\rangle\eqno (4)$$
where $M$ is chosen so that $\langle \zeta |\zeta\rangle=1$, so long as a
positive weight $k$ exists such that when integrated over the plane
$(\zeta=\xi
+i\eta)$
$$\one = \int |\zeta\rangle\langle \zeta|\, k (|\zeta|^2)\, d\xi\, d\eta
/\pi \, .\eqno(5)$$
Evidently $e^{-i{\cal H}_0t} |\zeta\rangle = |e^{-i\omega t}\zeta\rangle$
holds
for all $\zeta\in \ic$ for this alternative set of states just as well.  To
ensure (5) it suffices to have
$$\eqalign{\rho_n &=\int_0^\infty u^n \rho(u)\, du\, ,\cr
\rho(u)&\equiv M^2(u)\, k(u)\, ,\cr}\eqno(6)$$
and to generate such examples it is easiest to choose $\rho$ first.$^4$  As
one
example of such a set of alternative coherent states for the harmonic
oscillator we offer $\rho(u)=e^{-\sqrt{u}}/2$, with $\rho_n =(2n+1)!$ and
$M^2(u)=\sqrt{u}/\sinh\sqrt{u}$.  Do not look for a transitively acting group
 or
one up to a factor [as in (1)] that defines the states $|\zeta\rangle$ as
unitary
tranformations of a fiducial vector; there is no such group.$^5$  Such states
 are
generally not minimal uncertainty states, of course, but minimal uncertainty
states by themselves do not ensure temporal stability as illustrated by the
example of squeezed states.

\vskip.15truein
\vbox{\noindent{\bf Covering-space formulation}

As one further generalization we wish to extend the polar coordinates
$r,\theta~(z\equiv re^{i\theta})$, where $0\leq r<\infty$, $-\pi <\theta\leq
\pi$, to their covering space, namely the domain $0\leq r<\infty$, $-\infty
<\theta <\infty$.  To this end we introduce}
$$|r,\theta\rangle \equiv M(r^2) \sum_{n=0}^\infty
(r^ne^{in\theta}/\sqrt{\rho_n}\, )|n\rangle\eqno(7)$$
and a measure $\nu(r,\theta)$ defined by
$$\int F(r,\theta)\, d\nu(r,\theta)\equiv \lim_{\Theta\to\infty} {1\over 2
\Theta}
\int_0^\infty dr^2 k(r^2) \int_{-\Theta}^\Theta d\theta \, F(r,\theta).
\eqno(8)$$
It follows that the states $|r,\theta\rangle$ are continuous in $r$ and
 $\theta$,
and admit the resolution of unity
$$\one = \int |r,\theta\rangle\langle r,\theta|\, d\nu (r,\theta).\eqno(9)
$$
Evidently $\exp (-i {\cal H}_0 t)\, |r,\theta\rangle = |r,\theta-\omega
t\rangle$, which is how temporal stability appears in the present notation.

\vskip.15truein
\noindent{\bf Relaxing the functional dependence}

Prior to considering coherent states for the hydrogen atom, we first analyze a
simpler, yet related example.  In particular, we consider a single degree of
freedom system with a Hamiltonian given by ${\cal H}_1\equiv -\omega/(N+1)^2$,
namely one with eigenvalues $E_n\equiv -\omega/(n+1)^2$, $n=0,1,2,\ldots$.  As
coherent states for this example we choose ($0\leq s < \infty$, $-\infty <
 \gamma
<\infty$)
$$|s,\gamma\rangle\equiv M(s^2) \sum_{n=0}^\infty (s^n
e^{i\gamma/(n+1)^2}/\sqrt{\rho_n}\,) |n\rangle\, ,\eqno(10)$$
where each state $|n\rangle$ obeys ${\cal H}_1 |n\rangle = E_n|n\rangle$.
Clearly, the states $|s,\gamma\rangle$ are continuous and fulfill
$$\one=\int |s,\gamma\rangle \langle s,\gamma|\, d\nu (s,\gamma)\,
.\eqno(11)$$
It is trivial to observe that
$$ e^{-i{\cal H}_1t} |s,\gamma\rangle = |s, \gamma +\omega t\rangle\,
,\eqno(12)$$
establishing that the states in question exhibit temporal stability for ${\cal
H}_1$.  The states defined by (10) apply to a wide variety of choices for
$\rho(u)$ and its moments $\rho_n, n=0,1,2,\ldots$ .  Perhaps one of the
 simplest
examples of such coherent states arises when we choose $\rho(u)= e^{-u}$,
 $\rho_n
=n!$, and $M(s^2) =\exp (-s^2/2)$, in which case
$$|s,\gamma\rangle = e^{-s^2/ 2} \sum_{n=0}^\infty (s^n
e^{i\gamma/(n+1)^2}/\sqrt{n!}\, ) |n\rangle \eqno(13)$$
and
$$\one =\lim_{\Gamma\to\infty} {1\over \Gamma} \int_0^\infty
\int_{-\Gamma}^\Gamma |s,\gamma\rangle\langle s,\gamma|\, s\, ds\, d\gamma\,
.\eqno(14)$$

\vskip.15truein
\noindent{\bf Hydrogen-atom coherent states}

We now finally turn to the (bound state part of the) hydrogen atom.$^6$  We
characterize this example by a Hamiltonian ${\cal H}$ with spectrum $E_n=
- -\omega/(n+1)^2$, $n=0,1,2,\ldots$, $\omega=me^4/2$, and with a degeneracy
of
each level given by $(n+1)^2$, which in turn is spanned, for example, by
 standard
angular momentum states $|\ell m\rangle$, $0\leq \ell\leq n$, $-\ell \leq
m\leq
\ell$, as usual.  {\bf N}.{\bf B}.!  {\it In the present usage}\/,
$n=0,1,2,\ldots$, {\it while the standard usage for the hydrogen atom is}
$n=1,2,3,\ldots$ {\it for the principal quantum number}.  Thus the traditional
hydrogen atom bound state $|n \ell m \rangle$ becomes $|n + 1 \ell m\rangle$.
To accomodate the angular momentum states we introduce suitable hydrogen-atom
adapted angular-momentum coherent states$^{3,7}$
$$\eqalign{|n, \overline{\Omega}\rangle &\equiv \sum_{\ell=0}^n \sum_{m =
- -\ell}^\ell \left[{(2\ell)!\over (\ell+m)!(\ell-m)!}\right]^{1\over 2}\cr
&\quad\times \left( \sin {\overline\theta\over 2}\right)^{\ell-m}\! \left(
\cos
{\overline\theta\over 2}\right)^{\ell+m}\! \ e^{-i(m\overline\phi + \ell
\overline\psi)} |n + 1 \ell m\rangle \sqrt{2\ell+1}\cr}\eqno(15)$$
These hydrogen-atom adapted angular-momentum coherent states satisfy
$$\int |n, \overline\Omega \rangle\langle n, \overline\Omega |
\sin \overline \theta \, d \overline \theta \, d \overline \phi \, d \overline
\psi /8\pi^2 = \sum_{\ell =0}^ n \sum_{m = -\ell}^\ell |n + 1 \ell m\rangle
\langle n + 1 \ell m | = \one_n , \eqno(16)$$
which for each $n\geq 0$ is the unit operator when acting in an angular
 momentum
subspace in which $0\leq \ell \leq n$.  Observer, for all $n\geq 1$, that the
subspace in question carries a {\it reducible} representation of the rotation
group with a total dimensionality of $\sum_{\ell=0}^n (2\ell +1) = (n+1)^2$.

The appropriate coherent states for the hydrogen atom are then given by
$$|s,\gamma,\overline\Omega\rangle \equiv M(s^2)\sum_{n=0}^\infty (s^n
e^{i\gamma/(n+1)^2} /\sqrt{\rho_n}\, ) |n,\overline \Omega \rangle \, .
\eqno(17)$$
The coherent states in question are evidently continuous and furthermore
 satisfy
($d\overline\Omega \equiv \sin \overline \theta \, d \overline
\theta \, d \overline \phi \, d \overline \psi /8\pi^2$)
$$\one = \int |s,\gamma,\overline\Omega\rangle \langle s,\gamma, \overline
\Omega
|\, d\nu(s,\gamma)\, d\overline\Omega\, .\eqno(18)$$
We extend the definition of the number operator $N$ so that
$$N|n, \overline\Omega\rangle = n|n,\overline\Omega\rangle \, ,\eqno(19)$$
in which case the hydrogen-atom Hamiltonian reads ${\cal H} =-\omega
/(N+1)^2$,
$\omega=me^4/2$.  It follows that
$$e^{-i{\cal H} t} |s,\gamma, \overline\Omega\rangle = |s, \gamma +\omega t,
\overline\Omega\rangle \eqno(20)$$
demonstrating that the states in question have temporal stability.  Thus we
 have
established our goal of exhibiting coherent states with the required
continuity
and resolution of unity for the hydrogen atom, and which also exhibit temporal
stability.  Moreover, this goal has been realized for a multitude of possible
coherent-state sets based on various choices of the weight $\rho(u)$.

Finally, we turn our attention to exhibiting the hydrogen-atom coherent
states---at least as much as possible---by way of their configuration-space
representation.  Recall the standard spherical-coordinate representation of
hydrogen-atom eigenstates $|n+1\ell m\rangle$ given by$^8$
$$\langle r\theta\phi| n+1 \ell m\rangle = u_{n+1}^\ell (r) Y_{\ell m}
(\theta,
\phi)\, .\eqno(21)$$
Here $u_{n+1}^\ell$ denotes the usual radial hydrogen-atom eigenfunctions,
while
$Y_{\ell m}$ are the standard angular momentum eigenfunctions.  In turn,
$$\eqalignno{\langle r\theta\phi| n,\overline\Omega\rangle& \equiv
\sum_{\ell=0}^n u_{n+1}^\ell (r) \sum_{m=-\ell}^\ell \left [ {(2\ell)!\over
(\ell+m)!(\ell-m)!}\right ]^{1\over 2}
\left( \sin {\overline\theta\over 2}\right)^{\ell-m} \left( \cos
{\overline\theta\over 2}\right)^{\ell+m}\cr
\noalign{\vskip 15pt}
&\quad \times e^{-i(m\overline\phi + \ell\overline\psi)} Y_{\ell m} (\theta,
\phi)
\sqrt{2\ell +1}\, .&(22)\cr}$$
Finally
$$\eqalignno{\langle r\theta\phi| s,\gamma, \overline\Omega\rangle & = M(s^2)
\sum_{n=0}^\infty (s^n e^{i\gamma/(n+1)^2} /\sqrt{\rho_n}\, )\cr
&\quad \times \sum_{\ell=0}^n u_{n+1}^\ell (r) \sum_{m=-\ell}^\ell \left [
{(2\ell)!\over (\ell+m)!(\ell-m)!}\right ]^{1\over 2}
\left( \sin {\overline\theta\over 2}\right)^{\ell-m} \left( \cos
{\overline\theta\over 2}\right)^{\ell+m}\cr
\noalign{\vskip15pt}
&\quad \times e^{-i(m\overline\phi + \ell \overline\psi)} Y_{\ell m}
(\theta,\phi) \sqrt{2\ell +1}\, .&(23)\cr}$$
Furthermore, in units where $\omega=1$, the radial eigenfunctions are given by
$$u_{n+1}^\ell = N_{n+1}^\ell [2r/(n+1)]^\ell F(-n+\ell, 2\ell +2, 2r/(n+1))
e^{-r/(n+1)}\, ,\eqno(24)$$
where
$$N_{n+1}^\ell\equiv {1\over (2\ell +1)!} \sqrt{(n+\ell+1)!\over 2(n+1)
(n-\ell)!}
\left ({2\over n+1}\right)^{3\over 2}\eqno(25)$$
and
$$\eqalign{F&(-n+\ell, 2\ell+2,z)\cr
&= 1+{(\ell-n)\over (2\ell+2)} {z\over 1!} +{(\ell-n )(\ell-n+1)\over
(2\ell+2)(2\ell+3)} {z^2\over 2!} + {(\ell-n)(\ell-n+1)(\ell-n+2)\over
(2\ell+2)(2\ell+3)(2\ell+4)} {z^3\over 3!}+\cdots\cr} \eqno(26)$$
which has a last nonvanishing coefficient for $z^{n-\ell}$.

Armed with this coordinate-space representation of the coherent states one
may,
if desired, begin to choose an ``optimal'' weight function $\rho$, e.g., by
minimizing the uncertainty product for $\langle (r-\langle r\rangle)^2\rangle
\langle( p_r-\langle p_r\rangle)^2\rangle$ for given values of $s$ and
 $\gamma$,
etc.  Since different problems may well require different, problem-specific
optimizations, we shall not pursue this question further.  Instead we conclude
by making explicit one example of
hydrogen-atom coherent states, namely, those with $\rho_n =n!$, $n\geq 0$.  In
that case
$$\eqalign{\langle r\theta\phi| s,\gamma, \overline\Omega\rangle & =
e^{-s^2/2}
\sum_{n=0}^\infty (s^n e^{i\gamma/(n+1)^2} /\sqrt{n!}\, )\cr
&\quad \times \sum_{\ell=0}^n u_{n+1}^\ell (r) \sum_{m=-\ell}^\ell \left [
{(2\ell)!\over (\ell+m)!(\ell-m)!}\right ]^{1\over 2}
\left( \sin {\overline\theta\over 2}\right)^{\ell-m} \left( \cos
{\overline\theta\over 2}\right)^{\ell+m}\cr
&\quad \times e^{-i(m\overline\phi + \ell \overline\psi)} Y_{\ell m}
(\theta,\phi) \sqrt{2\ell +1}\, .\cr}\eqno(27)$$

\noindent{\bf Summary}

In (23), and more specifically in (27), we propose a set of coherent
states appropriate to the bound state portion of the hydrogen atom.
In the definition adopted, these states involve 5 real parameters,
namely $s$, $\gamma$, and $\overline\Omega = (\overline \theta,
\overline \phi, \overline \psi)$. \ The coherent states (i)~are
continuous in these 5 parameters, (ii)~admit a resolution of unity,
(18), as a positive integral over one-dimensional projection
operators, and (iii)~evolve into one another under time evolution
with the Hamiltonian of the bound-state hydrogen atom, (20). \
There is some arbitrariness in the definition in (23) that would
permit optimization of some additional feature(s) of the coherent
states.
\vfill\eject

\centerline{\bf References}
\item{1.} E. Schr\"odinger, Naturwissenshaften {\bf 14}, 644 (1926).
\item{2.}R.J. Glauber, Phys. Rev. Letters {\bf 10}, 84 (1963); J.R. Klauder
and
E.C.G. Sudarshan, {\it Fundamentals of Quantum Optics} (W.A. Benjamin,
 New York,
1968), Chap. 7.
\item{3.}J.R. Klauder and B.-S. Skagerstam, {\it Coherent States} (World
Scientific, Singapore, 1985).
\item{4.} We assume $\rho$ is chosen so that all moments exist and that the
sum
in (4) converges strongly for all $\zeta\in \ic$.
\item{5.}J.R. Klauder, Mod. Phys. Letters A{\bf 8}, 1735 (1993); J.R. Klauder,
Annals of Physics {\bf 237}, 147 (1995).
\item{6.} Some examples of previous studies of coherent states (variously
defined!) for the hydrogen atom include:  L.S. Brown, Am. J. Phys. {\bf 41},
 525
(1973); J. Mostowski, Lett. Math. Phys. {\bf 2}, 1 (1977); J.C. Gay
{\it et al.}, Phys. Rev. A {\bf 39}, 6587 (1989); M. Nauenberg, Phys.
Rev. A {\bf 40}, 1133 (1989); Z.D. Gaeta and C.R. Stroud, Jr., Phys. Rev. A
 {\bf
42}, 6308 (1990); J.A. Yeazell and C.R. Stroud, Jr., Phys. Rev. A {\bf 43},
 5153
(1991); M. Nauenberg, in {\it Coherent States:  Past, Present, and Future},
 eds.
D.H. Feng, J.R. Klauder, and M.R. Strayer (World Scientific, Singapore, 1994),
p. 345; I. Zlatev, W.-M. Zhang, and D.H. Feng, Phys. Rev. {\bf 50}, R1973
(1994);
R. Bluhm, V.A. Kostelecky, and B. Tudose, Colby College
Preprint 95--07.
\item{7.}A.M. Perelomov, {\it Generalized Coherent States and Their
 Applications}
(Springer-Verlag, Berlin, 1986).
\item{8.} A.S. Davydov, {\it Quantum Mechanics} (Addison-Wesley, Reading, MA,
1968).
\end